\documentclass[
 reprint,
 groupedaddress,
superscriptaddress,
 amsmath,amssymb,
 aps,
]{revtex4-2}

\usepackage{graphicx}
\usepackage{dcolumn}
\usepackage{bm}
\usepackage{hyperref}
\usepackage{xcolor}
\usepackage{hhline} 
\usepackage[normalem]{ulem}

\begin{document}

\preprint{APS/123-QED}

\title{Second quantization of anyons and spin–anyon duality}

\author{Priyanshi Bhasin}
\affiliation{Department of Physics, Indian Institute of Science, Bangaluru, 560012, India}
\author{Diptiman Sen}
\affiliation{Center for High Energy Physics, Indian Institute of Science, Bengaluru 560012, India}
\author{Tanmoy Das}
\affiliation{Department of Physics, Indian Institute of Science, Bangaluru, 560012, India}

\begin{abstract}
Anyons exhibit a non-trivial interplay between local exclusion rules and non-local braiding and exchange phases, making a consistent commutation algebra and second-quantized formulation challenging.
We develop an algebraic framework for Abelian anyons in one dimension with statistical phase $\theta=\pi/N$ that enforces a finite on‑site occupancy of $N-1$ anyons with the exchange phase $\theta$ between different sites. Moreover, we introduce an exact Jordan–Wigner duality between $\pi/3$ anyons and spin-1 operators, allowing us to map a tight‑binding anyon model to an XY‑like spin‑1 model. The model exhibits anyon‑density–dependent flux, incompressible or gapless regions, and critical points with level crossings that appear as discontinuities in ground‑state currents, momenta, fidelities, and correlation functions. Our second-quantization formalism establishes a novel spin–anyon duality, offering a conceptually new route to realize anyons from spin Hamiltonians and to engineer corresponding device architectures. 
\end{abstract}

\maketitle

\section{Introduction}

Anyons - exotic quasiparticle excitations emerging in fractional quantum Hall (FQH) like settings\cite{leinaas1977, goldin1981representations, wilczek1982quantum, wilczek1982magnetic, tsui1982two, laughlin1983anomalous, halperin1984, arovas1984fractional, wen1990chiral, nayak2008non} 
or lattice gauge models~\cite{kundu1999exact, batchelor2006one, keilmann2011statistically, greschner2015anyon, tang2015ground, cardarelli2016engineering, strater2016floquet, agarwala2019statistics, aditya2021bosonization, brollo2022two} - exhibit distinctive interplay between locality and non‑locality. Locally, they obey exclusion rules that must bound the on‑site occupancy, while non‑locally they acquire exchange and braiding phases, both of which are set by a statistical angle $\theta$.\cite{halperin1984, haldane1991fractional, polychronakos1996probabilities, arovas1984fractional, bartolomei2020fractional} Despite progress in continuum and first‑quantized approaches,\cite{leinaas1977, wilczek1982quantum, wilczek1982magnetic,batchelor2006one}, a closed operator algebra and its second-quantized particle‑based formalism that simultaneously enforces finite local occupancy and exchange phases have remained elusive.

Recent efforts to engineer anyon-like algebraic structures start from conventional bosonic or fermionic systems, introducing density-dependent gauge fields so that hardcore bosons acquire tunable exchange phases and emulate anyonic statistics~\cite{rabello1995gauge, rabello19961d, kundu1999exact, keilmann2011statistically, Greiner2024}. Subsequent progress has been partial, achieving tight-binding models for anyons that capture exchange statistics~\cite{roushan2017chiral, clark2018observation, lienhard2020realization, yao2022domain, frolian2022realizing}, but not the local exclusion statistics. 

Traditional algebraic approaches based on graded algebras among ladder and number operators do not fully account for these requirements, as will be demonstrated later. Modified algebraic structures, including the so-called quantum algebra~\cite{arik1976hilbert,biedenharn1989quantum,zagier1992realizability,greenberg1990example,fivel1990interpolation,greenberg1991particles,bozejko1991example,chaturvedi1991generalized} or $q$-deformed quantum groups~\cite{macfarlane1989q,biedenharn1989quantum,yan1990q,chaichian1993statistics,narayana2006interpolating}, have limited success in single-particle descriptions, with no significant progress in their many-body description. 

Here we introduce an algebraic construction for Abelian anyons in one dimension with $\theta=\pi/N$ that addresses both requirements. We recognize that the deformed commutation between creation and annihilation operators: $\hat{C}=bb^{\dagger}-e^{i\theta}b^{\dagger}b$ is intrinsically non‑Hermitian. We define a Hermitian number operator $\hat{N}$ not as $b^{\dagger}b$ but through the deformed algebra as  $\hat{C}=e^{-i\hat{N}\theta}$~\cite{bhasin2025hermitian}. 
The resulting local spectrum consists of uniformly graded eigenstates of $\hat{N}$; while the ladder amplitudes $b^{\dagger}b$ involve Chebyshev‑type polynomials, limiting the Fock space dimension at $n=0, 1, ..., N-1$~\cite{bhasin2025hermitian}. In this way, the algebra simultaneously imposes a finite on‑site occupancy of $N-1$ anyons and ensures the off‑site exchange relations $b_i b_j = e^{-{\rm i}\theta} b_j b_i$, for $i>j$. The bosonic limit appears smoothly for $N\rightarrow \infty$, while special cases such as semions are recovered for any even $N$.  

We propose a ring of interacting tight-binding anyons that can be exactly derived from a modified $XY$ spin model with a site- and spin-dependent gauge field and a uniform Zeeman field. We analyze the mapped model by exact diagonalization on finite rings, with periodic boundary condition (PBC). The flux threading the ring becomes anyon density dependent, thereby inducing incompressibility or energy gap between distinct anyon-number sectors. The ground state generically carries a finite total momentum, and supports a persistent current that depends sensitively on the anyon density and hopping amplitude. As parameters are tuned, the spectrum exhibits level crossings between the ground and first excited states; these crossings are accompanied by discontinuities in the persistent current and by sharp features in the fidelity and correlation functions in the ground state. Incompressible regions occur at specific fillings, whereas gapless points mark transitions between momentum sectors.

The rest of the paper is structured as follows. In Sec.~\ref{Sec:Quantization}, we develop the second quantization algebra of anyons and the corresponding single-particle and many-body Fock states. We discuss a one-dimensional tight-binding anyon model and the Jordan-Wigner duality to spin-1 operators in Sec.~\ref{Sec:lattice_model}. We present numerical results obtained via exact diagonalization of the lattice model in Sec.~\ref{Sec:Results}, including the energy gap, ground-state current, ground-state fidelity, and spin-spin correlation functions. Finally, in Sec.~\ref{Sec:Discussions} we discuss the advantages and limitations of the model and its prospects in engineering anyons through spin models. In the Appendix, we give further details of the theory and numerical results, including consistency checks of exchange, braiding statistics, hardcore anyon limit dual to constrained spin-1/2 fermion model, system size dependence of results, and others.

\section{Algebraic Quantization of Anyons}\label{Sec:Quantization}

Let $b_i$ and $b_i^\dagger$ be the anyonic annihilation and creation operators, respectively, at a lattice site $i$, and $\hat{N}_i\ne b_i^{\dagger}b_i$ be a (diagonal) number operator obeying the following algebraic relations \footnote{The prescribed algebra stems from a more general consideration of $b_i b_j^\dagger -e^{i\theta} b_j^\dagger b_i = C_i\delta_{ij}$ where $C_i\ne C_i^{\dagger}\ne \mathbb{I}$. Using polar decomposition, we can write $C_i=U_i|C_i|$, where $|C_i|=\sqrt{C_i^{\dagger}C_i}$ is invertible, and $U_i$ is a unitary operator. Writing $U_i=e^{-i\theta \hat{N}_i}$ where $N_i$ is a Hermitian operator and rescaling $b_i\rightarrow |C_i|^{-1/2}b_i$, we arrive at Eq.~\eqref{Eq: Anyon bbdaggerN}.}:
\begin{subequations}\label{Eq: Anyon algebra}
\begin{align}
    b_i b_j^\dagger -e^{i\theta} b_j^\dagger b_i &= e^{-{\rm i} \hat{N}_i \theta} \delta_{ij}, \label{Eq: Anyon bbdaggerN}\\
    b_i b_j &= e^{-{\rm i}\theta} b_j b_i, \label{Eq: Anyon bb}\\
    b_i^\dagger b_j^\dagger &= e^{-{\rm i}\theta} b_j^\dagger b_i^\dagger, \label{Eq: Anyon bdaggerbdagger}  \\
    (b_i)^N &= (b_i^\dagger)^N = 0.    \label{Eq: Anyon exclusion}
\end{align}
\end{subequations}
Given that clockwise and anticlockwise exchanges are inequivalent, we adopt an ordering $i>j$ in Eqs.~\eqref{Eq: Anyon bbdaggerN}-\eqref{Eq: Anyon bdaggerbdagger}. $\hat{N}_i$ is a Hermitian operator whose eigenstates provide the grading of the states as $\hat{N}_i |n_i \rangle =n_i |n_i \rangle$, $\forall i$ where $n_i \in \mathbb{Z}_{+}$. The appearance of the phase $-\theta$ on the right side of the exchange equations~\eqref{Eq: Anyon bb}and~\eqref{Eq: Anyon bdaggerbdagger} is fixed by the requirement that the local operator $b_i^\dagger b_i$ is a Hermitian operator. (We occasionally use a hat symbol over some, but not all, operators where the meaning can be obscured). 

For each $i$, we find interestingly that $\hat{U}=b^{\dagger}b = \sin{(\hat{N}\theta)}/\sin \theta$, whose eigenvalues are the Chebyshev polynomials of the second kind, $U_{n-1}(\cos\theta)$,  
\begin{equation}\label{Eq: Anyon betan}
    U_{n-1}(\cos\theta) = \frac{\sin (n\theta)}{\sin(\theta)}:=\beta_n.
\end{equation}
Here, both $\beta_0=\beta_N=0$, restricting the states to an $N$-dimensional Fock space for $n=0, 1, 2,\cdots, N-1$. Additionally, we observe a degeneracy at $\beta_n=\beta_{N-n}$. 
The ladder actions are $b|n\rangle = \sqrt{\beta_n}|n-1\rangle, \,
    b^\dagger|n\rangle = \sqrt{\beta_{n+1}}|n+1\rangle$, with $b^\dagger b |n\rangle = \beta_n|n\rangle$ and $b b^\dagger |n\rangle = \beta_{n+1}|n\rangle$, and the normalized Fock states are
\begin{equation}\label{Eq: Anyon n in b dagger}
    |n\rangle = \frac{\left(b^\dagger \right)^{n}}{{\mathcal{N}_n}}|0\rangle, 
\end{equation}
with $\mathcal{N}_n=\sqrt{\beta_1\beta_2\cdots \beta_n}$. The states $|0\rangle$ and $|N-1\rangle$ turn out to be exceptional points of $b$ and $b^{\dagger}$, respectively, and hence are annihilated by them~\cite{bhasin2025hermitian}. We note that the states $|n\rangle$ are uniformly graded between the two null states, where $\beta_n(\theta)$ takes fractional values according to $\theta$. This makes the formalism distinct, and helps capture the anyon exclusion principle given in Eq.~\eqref{Eq: Anyon exclusion}, while the remaining equations (Eqs.~\eqref{Eq: Anyon bbdaggerN}-\eqref{Eq: Anyon bdaggerbdagger}) give the exchange and braiding statistics of Abelian anyons.

Our construction coincides with the bosonic case for $N\rightarrow \infty$, and $n<<N$ such that $\beta_n\approx n$. The $N=2$ condition leads to semions which have an exchange phase of $\theta = \pi/2$, and appear in chiral spin liquid states~\cite{kalmeyer1987equivalence,wen1989chiral,gorohovsky2015chiral,ferraz2019spin} 
For $N>2$ with any even $N$, the quasiparticles are still Abelian anyons~\cite{wen2004quantum}
and the present theory is applicable.

Given that the anyonic exchange between neighboring sites depends on the direction of exchange, the many-body state construction also requires an ordering convention. Following the same convention as in Eq.~\eqref{Eq: Anyon algebra}, with an $N$ dimensional Fock space per site, we construct a product state in a linear chain of $L$ sites as
\begin{equation}\label{Eq: Anyon many body state} 
   |n_1, n_2, \cdots ,  n_L\rangle = \frac{\left(b_1^\dagger \right)^{n_1}}{{\mathcal{N}_{n_1}}} \frac{\left(b_2^\dagger \right)^{n_2}}{{\mathcal{N}_{n_2}}}\cdots \frac{\left(b_L^\dagger \right)^{n_L}}{{\mathcal{N}_{n_L}}}|{0}\rangle, 
\end{equation}
where $n_i$ denotes the occupation number at site $i$ , with $M=\sum_{i=1}^Ln_i$ is the total anyon number, and $|0\rangle$ is the global vacuum at all sites. This many-body state incorporates both the on-site exclusion principle and the non-local exchange statistics. A PBC with site $L+1$ equal to site 1 is not easy to implement for the states, and we will instead impose it in the Hamiltonian below.  

Before ending this section, we note that
it is possible to define a different model of anyons where
there are again $N$ states at each site, but $b_i$ and
$b_i^\dagger$ act in a {\it cyclical} way. Namely, $b_i^\dagger$ acting on the state $| N-1 \rangle$ takes us to the state $| 0 \rangle$ (instead of
annihilating) and $b_i$ acting on $| 0 \rangle$ takes us back to $| N-1 \rangle$. This would require that the anyon number can change modulo $N$, with the change being accommodated by
appropriate modifications of the sea of electrons.
The algebra required to describe such a model is
clearly very different from the one given in 
Eqs.~\eqref{Eq: Anyon algebra}.
We will not consider this alternative model further in this paper.

\section{Lattice model} \label{Sec:lattice_model}

\begin{figure}[t]
    \centering    \includegraphics[width=0.95\columnwidth]{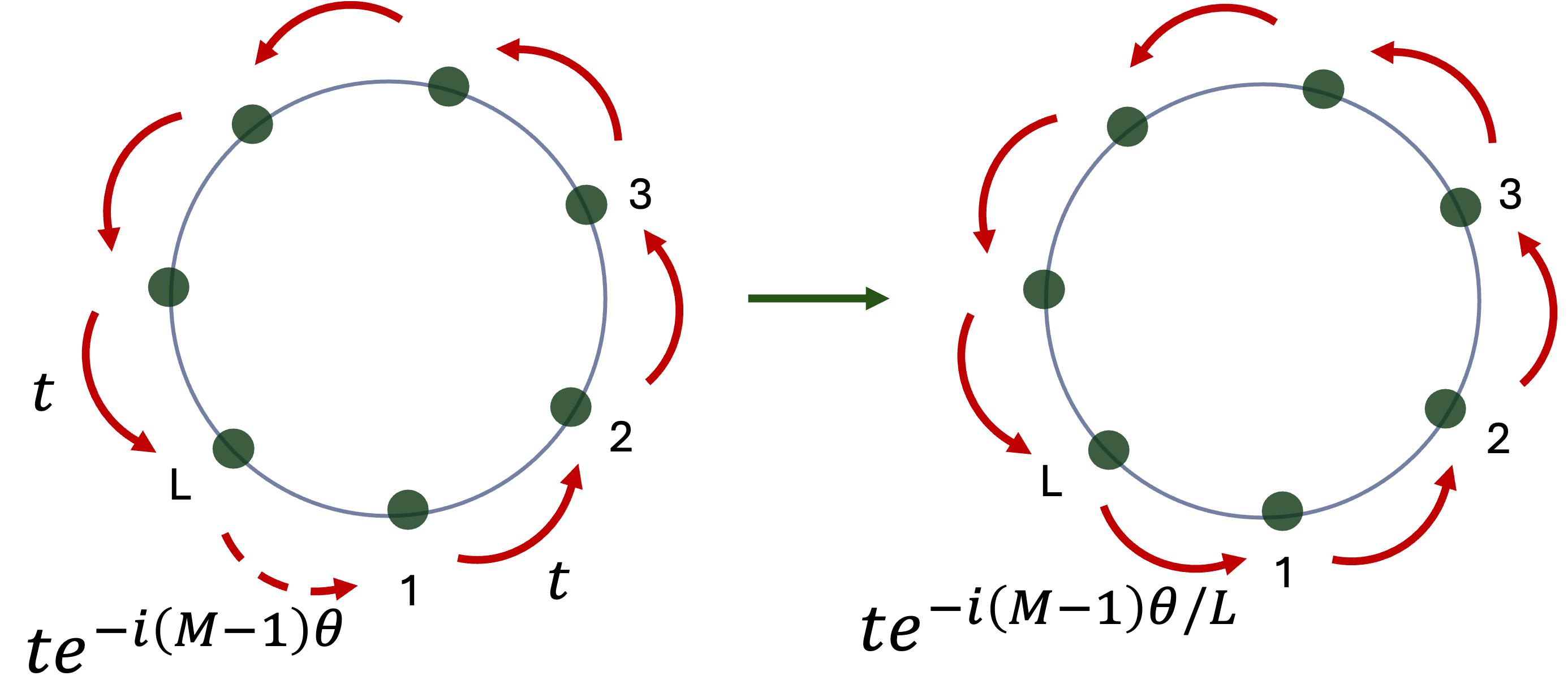}
    \caption{\textbf{Schematic of an anyon chain with PBC.} The PBC is imposed by a phase $(M-1)\theta$ which can be either inserted in the Hamiltonian or many-body state. Here we can impose the phase either at a given site (left panel), or distribute it equally on each bond as $(M-1)\theta/L$ (right panel). The advantage of the second approach, as done here, is that it makes the total momentum a good quantum number. 
    }
    \label{Fig:Ring hopping}
\end{figure}

Next, we consider a 1D lattice model for anyons with nearest-neighbor tight-binding hopping term ($H_1$) and an on-site Hubbard interaction term ($H_2$) with PBC:
\begin{equation}\label{Eq: Ham hopping PBC}
    H_1 = t \sum_{j=1}^L   
    \left[e^{i\rho} b_j^\dagger b_{j+1} + e^{-i\rho} b_{j+1}^\dagger b_j \right],
\end{equation}
where $\rho = (M-1)\theta/L$, with $M$ denoting the total number of anyons and $\nu = M/L$ being the anyon density. For later convenience, we define $\nu_{-1}= (M-1)/L$ as the anyon density with one less anyon in the system. Consequently, $\rho=\nu_{-1}\theta$ can be interpreted as the corresponding flux density with one lesser anyon. The origin of the phase $\rho$ appearing in Eq.~\eqref{Eq: Ham hopping PBC} is to impose PBC, as shown in Appendix~\ref{Appendix: Anyon tight binding hamiltonian} (refer Fig.~\ref{Fig:Ring hopping} for a schematic demonstration).

The on-site Hubbard interaction is incorporated in terms of the operator $\hat{N}_i$ as,
\begin{equation} \label{Eq: onsite hubbard}
    H_2 = \sum_{j=1}^L\left[ \frac{U}{2}\hat{N}_j(\hat{N}_j-1) + \mu \hat{N}_j \right],
\end{equation}
where $U$ is the Hubbard interaction strength and $\mu$ is the chemical potential. (We note that the on-site energy is incorporated by using the operator $\hat{N}_i$, instead of  $b^\dagger_i b_i$). For $\pi/3$ anyons with zero, single, and double occupancies in a single site, we associate the corresponding on-site energies to be $0$, $\Delta_1$, $\Delta_2$. This amounts to $U=\Delta_2-2\Delta_1$ and $\mu = \Delta_1$ in Eq.~\eqref{Eq: onsite hubbard}. For $\Delta_2 > 2\Delta_1$, the interaction is repulsive and penalizes double occupancy. 

\subsection{Spin-anyon duality}

We note that the commutation algebra for the on-site anyonic operators gives $[N_i,b_i]=-b_i$, $[N_i,b^\dagger_i]=b^\dagger_i$, $[b_i, b_{i}^{\dagger}]= \cos \left((2\hat{N}_i+1)\frac{\theta}{2}\right)/\cos (\frac{\theta}{2})$, which coincides with the algebra of spin-1 operators for $\theta=\pi/3$. In the rest of this paper, we will primarily consider the case 
$\theta=\pi/3$ and $N=3$ for the sake of simplicity. Then there are three states allowed at each lattice site which correspond
respectively to $0, 1$ and 2 anyons.
While spin operators commute between different sites, anyonic operators do not. Hence we introduce a non-local Jordan-Wigner string operator to develop an isomorphism between them as 
\begin{subequations}\label{Eq: Anyon JW}
    \begin{align}
        \hat{N}_i &= S^z_i + \mathbb{I}, \\
         b_i  &= \frac{1}{\sqrt{2}} e^{+{\rm i}\theta\sum_{j=1}^{i-1}\hat{N}_j} S^-_i,\\
        b^\dagger_i &= \frac{1}{\sqrt{2}} S^+_i  e^{-{\rm i}\theta\sum_{j=1}^{i-1}\hat{N}_j}.
    \end{align}
\end{subequations}
Here $\mathbb{I}$ is a $3\times 3$ identity matrix. 
For other $\theta\ne \pi/3$ cases, a modified Jordan-Wigner transformation can be found between the present anyon operator and another set of anyon-like operators (not spin) introduced in Appendix~\ref{Appendix: Anyon JW proof}.

In the spin-1 language, the Hamiltonian $H=H_1+H_2$ becomes,
\begin{align}\label{Eq: Anyon hamiltonian spin periodic}
    H &= \frac{1}{2} \sum_{j} \left[ 
    J_{ex}  S_{j}^{+} e^{{\rm i}S_j^z\theta} S_{j+1}^{-}
    + J_{ex}^* S_{j+1}^{+} e^{-{\rm i}S_j^z\theta} S_{j}^{-} 
    \right] \nonumber \\
    & \qquad + \sum_{j} \left[ \frac{U}{2} (S^z_j)^2 + h S^z_j   \right]+L\mu,
\end{align}
where $J_{ex}=te^{i\theta}\in\mathbb{C}$, $h=\Delta_2/2\in\mathbb{R}$ give the complex exchange and real Zeeman fields, and $U$ acts as the spin anisotropy energy.  Eq.~\eqref{Eq: Anyon hamiltonian spin periodic} is obtained from Eqs.~\eqref{Eq: Ham hopping PBC} and \eqref{Eq: onsite hubbard}, with Eq.~\eqref{Eq: Anyon JW} and after applying a global gauge transformation $\mathcal{U}=\prod_j e^{-i \nu_{-1}(j-1)S^z_j\theta}$. The Hamiltonian in Eq.~\eqref{Eq: Anyon hamiltonian spin periodic} is like an $XY$ model with occupancy-dependent exchange term, spin anisotropy, and a Zeeman term along the $z$-direction. Interestingly, while the anyonic Hamiltonian (Eqs.~\eqref{Eq: Ham hopping PBC} and \eqref{Eq: onsite hubbard}) is defined fully locally with a uniform gauge field $\rho$ at each bond, the same in the spin language acquires a site-dependent phase factor of $e^{i S_j^z \theta}$. Note that for both the forward and backward exchange terms in Eq.~\eqref{Eq: Anyon hamiltonian spin periodic}, the phase factor $e^{i S_j^z \theta}$ depends on the $S^z_j$ value of the site to the left, i.e., 
the $j$-th site. This breaks the spatial inversion symmetry, which is a manifestation of the directional asymmetry of the anyon algebra in Eq.~\eqref{Eq: Anyon algebra}.

In the next section, we will use Eq.~\eqref{Eq: Anyon hamiltonian spin periodic} to numerically study our system with 
a finite number of sites, denoted as $L$. We will assume PBC so that the sum over $j$ in the Hamiltonian
goes from $1$ to $L$, with $S_{L+1}^a \equiv S_1^a$ where $a= \pm, z$.

\section{Numerical results}\label{Sec:Results}

\begin{figure}[t]
    \centering    \includegraphics[width=0.95\columnwidth]{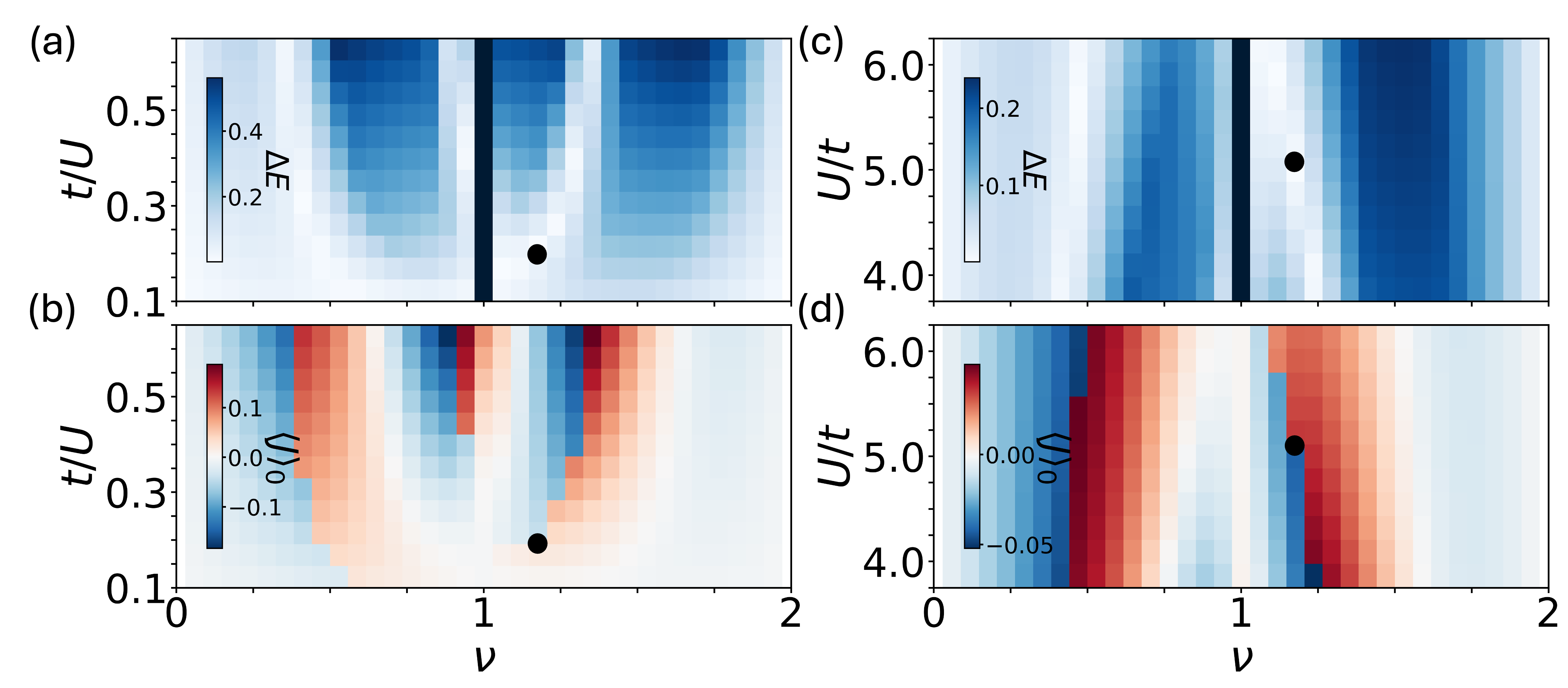}
    \caption{\textbf{Persistent current and energy gap as a function of filling fraction ($\nu$), hopping amplitude ($t$), and onsite interaction strength ($U$).} \textit{Left panel:} (a) Energy gap between the first excited state and the ground state $\Delta E= E_1 - E_0$ is plotted as a heatmap as a function of $\nu$ and $t/U$. The fixed parameters are $\Delta_1=1$, $\Delta_2=4$, and the system size is $L=17$.  White regions denote the suppression or closing of the energy gap.  
    The dark blue column at $\nu =1$ denotes a very large gap $\Delta E \sim \mathcal{O}(U)$. (b) The corresponding ground-state persistent current $\langle J \rangle_0$ is shown at the same parameter values. Red and blue regions indicate positive and negative currents, respectively, while the white region denotes zero current. $\langle J \rangle_0$ exhibits discontinuous sign reversals at intermediate fillings as $t/U$ is increased. The sign reversal points are correlated with the energy gap suppression regions in panels (a) and (b). 
  \textit{Right Panel:} (c,d) Same quantities plotted as a function of $U/t$ with $t=0.4$ and $\Delta_1=1$.  The black circles mark a representative level crossing point, where detailed analysis is done.
    }
    \label{Fig:Heatmap PC GS M vs t}
\end{figure}

Note that the anyon algebra in Eq.~\eqref{Eq: Anyon algebra} does not canonically transform to the same anyonic form in momentum space. Hence the Fourier transformation of Eqs.~\eqref{Eq: Ham hopping PBC}, \eqref{Eq: onsite hubbard} does not help here. We perform an exact diagonalization of  Eq.~\eqref{Eq: Anyon hamiltonian spin periodic} using the QuSpin software package. We mainly study the ground state and a few low-energy excited states as a function of $t$, $\Delta_{1,2}$, and $\nu$. We discuss the results for $L=17$, the overall qualitative conclusions remain invariant with system size which is discussed in Appendix~\ref{Appendix: System size invariance}.

The model is non-integrable, with two apparent global symmetries:  $U(1)$ and translational symmetries, rendering the total anyon number $M$ (i.e., total $S^z$ for $\theta=\pi/3$ ) and the total momentum $P$ to be conserved (but single particle momentum is not well defined).  The corresponding wavevectors are given by  $K = \frac{2\pi}{L}k$, where $k=0,1,2,\cdots ,L-1$ under PBC. In the numerical computations, these momenta are extracted by applying the global translation operator $ e^{ia{P}/\hbar}$ that translates all anyons/spins by one lattice constant $a$ (we will henceforth set $a=1$, and $\hbar=1$), 
\begin{equation}
    e^{i{P}} |n_1, n_2,\cdots ,n_i,\cdots ,n_L\rangle = |n_L, n_1,\cdots ,n_{i+1},\cdots ,n_1\rangle.
\end{equation}
The ground and low-energy excited states have finite momenta $P$, and finite current, presumably due to finite flux $\phi=\rho L$ in the lattice under PBC. In the hardcore limit $\Delta_2\rightarrow \infty$, the anyon model reduces to a hardcore boson or fermion or spin-1/2 model, where single-particle momentum and band dispersions become well-defined (See Appendix~\ref{Appendix: hard core}). 

We define the current operator $J$ from the continuity equation with the local density operator $\hat{N}_j$  to obtain (see Appendix~\ref{Appendix: Anyon PC} for details),
\begin{subequations}
    \begin{eqnarray}
    J &=& -\frac{it}{ L}\sum_{j=1}^L\left[ 
    e^{i\rho}b_j^\dagger b_{j+1}
    -
    e^{-i\rho} b_{j+1}^\dagger b_j
    \right] \label{Eq: Anyon pc bbdagger}\\
    &=&-\frac{i}{2 L}\sum_{j=1}^L\left[
    J_{ex}  S_{j}^{+} e^{iS_j^z\theta} S_{j+1}^{-}
    -
    J_{ex}^*  S_{j+1}^{+} e^{-iS_j^z\theta} S_{j}^{-}
    \right] \qquad \label{Eq: Anyon pc spsm}
\end{eqnarray}
\end{subequations}

\begin{figure}[t]
    \centering    \includegraphics[width=0.95\columnwidth]{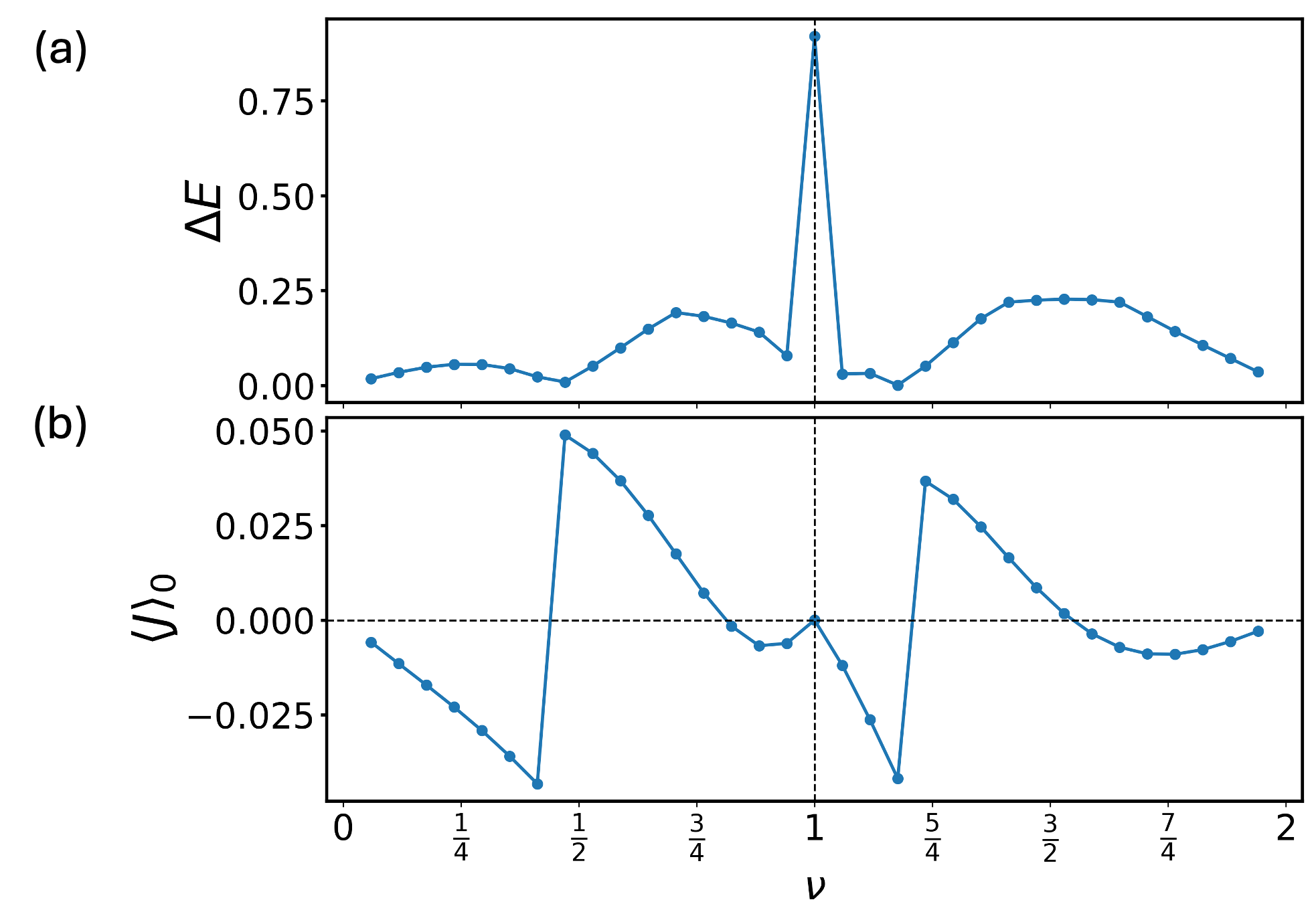}
    \caption{\textbf{Energy gap and persistent current.} 
    (a) Energy gap $\Delta$ (upper panel), and (b) the ground state current $\langle J \rangle_0$ (lower panel), plotted as a function of anyon number density $\nu$ , at fixed $t=0.4$, $\Delta_1=1$, and $\Delta_2=4$ (i.e., $U=2>0$). Dots represent computed data and the lines are guides to the eyes. The vertical dashed line marks half-filling ($M = L$).  
    The sign reversals in $\langle J \rangle_0$
    are associate with minima in $\Delta E$.}
    \label{Fig:PC 1st 2nd jump eg}
\end{figure}

\begin{figure}[t]
    \centering    \includegraphics[width=0.95\columnwidth]{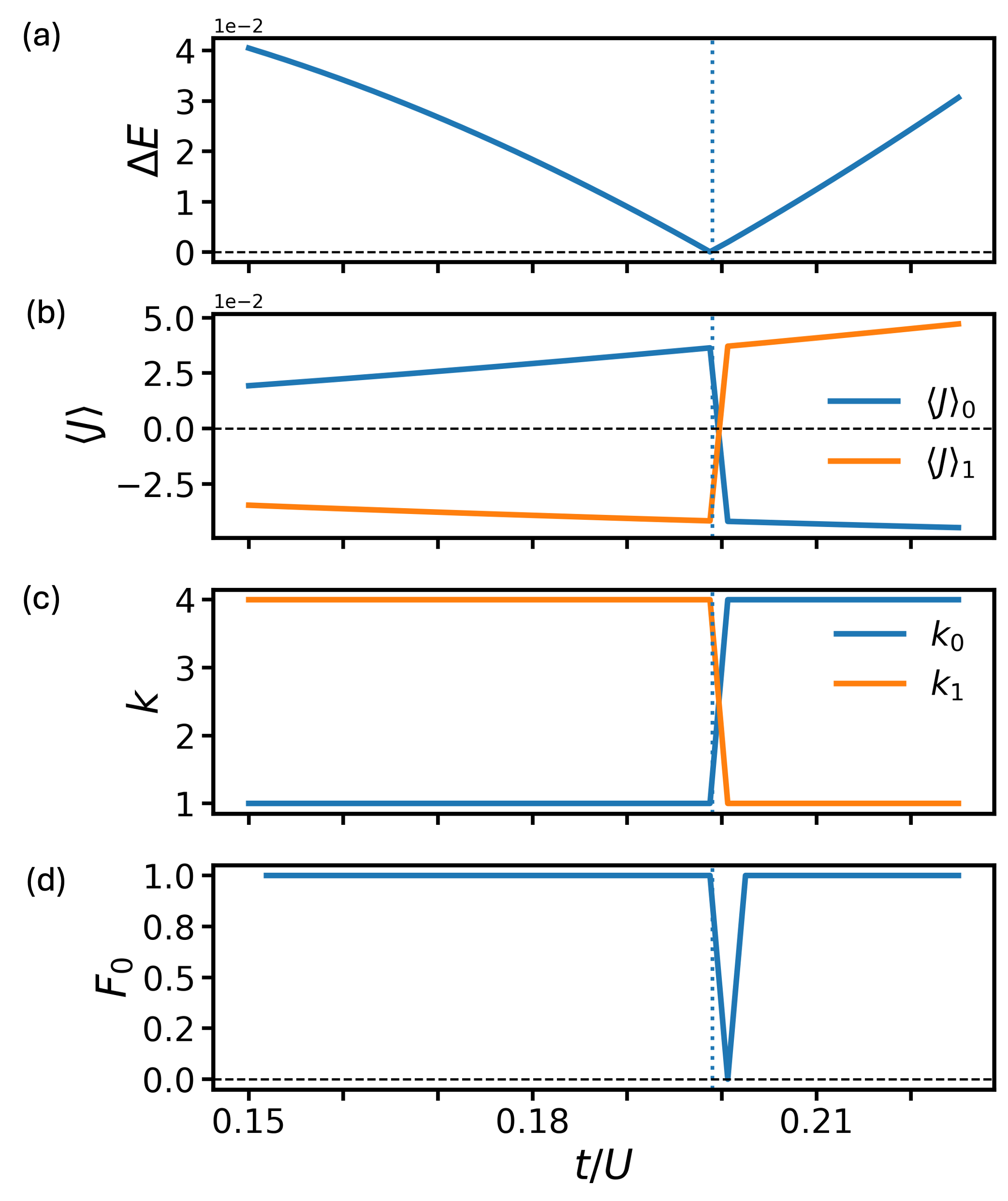}
    \caption{\textbf{Level crossing as a function of hopping amplitude.} 
    We show the results of (a) $\Delta E$ (b) ground state fidelity $F_0$, (c) current in ground/first excited states $\langle J \rangle_{0/1}$, and (d) momentum of the ground/first excited state ($k_{0/1}$), as a function of hopping amplitude ($t/U$) for $\nu = 20/17$, $\Delta_1=1$, and $\Delta_2=4$. Vertical dashed line marks the critical point $t_c=0.4$ where $\Delta E=0$, accompanied by a sharp dip in $F_0$, and a discontonuous jump in $\langle J \rangle $, and $k$.}
    \label{Fig:PC first jump}
\end{figure}

We study the Hamiltonian in Eq.~\eqref{Eq: Anyon hamiltonian spin periodic} using the exact diagonalization method for a system size of $L=17$. Fig.~\ref{Fig:Heatmap PC GS M vs t} shows a phase diagram of energy gap $\Delta E=E_1-E_0$ between first excited state and ground state, and the ground state value of the current $\langle J\rangle_0$ as a function of hopping $t$ and filling fraction $\nu$, at a representative value of  $\Delta_1=1$, and $\Delta_2=4$ (i.e., $U=2$). In Fig.~\ref{Fig:PC 1st 2nd jump eg}, we take a representative cut along $\nu$ for fixed $t=0.4$, while the same results along $t$ for fixed $\nu=20/17$ are shown in Fig.~\ref{Fig:PC first jump}. In all cases, we observe a coincidence of the level inversion, discontinuous jumps in $\langle J\rangle_0$. 

At half-filling ($\nu=1$) the energy gap $\Delta E$ becomes large, and subsequently the current vanishes, bit does not necessarily change sign, see Fig.~\ref{Fig:PC 1st 2nd jump eg}(a). On the other hand, $\langle J\rangle_0$ shows prominent jump and sign reversal on both sides of half-filling 
$\nu\sim 1/2$, and $\nu\sim 5/4$, see Fig.~\ref{Fig:PC 1st 2nd jump eg}(b). Both jumps are corroborated with a corresponding local minima in $\Delta E$, suggesting that the level inversion  causes the jump in current.

Since $t$ is a continuous variable, the level inversions and jumps in current are clearly resolved as a function of $t$ in Fig.~\ref{Fig:PC first jump}. At the critical value of $t_c=0.4$, the ground state and first excited states become exactly degenerate, with a level inversion across it. This is evidenced the sign reversal in $\langle J\rangle_{0/1}$ and the changes between the values of $k_{0/1}$. Note that the total momentum remains finite in both ground state and first excited states, and consequently persistent current remains finite in both states due to the loss of chirality.

Moreover, since $t$ is a continuous parameter, we can study the ground state fidelity as a function of $t$ to look for quantum phase transitions. The fidelity of a quantum state $|\phi\rangle$, defined as, $F(t+\delta t) \equiv |\langle\phi(t)|\phi(t+\delta t)\rangle|$, quantifies the overlap between the ground states $|\phi(t)\rangle$ of the Hamiltonian due to  an infinitesimal change in the parameter $t$. If $|\phi(t)\rangle$ and $|\phi(t+\delta t)\rangle$ become linearly independent, we have $F=0$; otherwise, it takes a finite value that rises up to $F=1$ when the two ground states are exactly the same~\cite{yang2025,tang2021}.  Our results in Fig.~\ref{Fig:PC first jump}(b) shows a jump in $F$ from $1$ to $0$ at $t_c$ where the ground state and the first excited state become exactly degenerate.

Fig.~\ref{Fig:Heatmap PC GS M vs t} consolidates these level inversions in the $M$ versus $t$ parameter space. We notice that despite the $\beta_n=\beta_{N-n}$ symmetry, the symmetry across half-filling is not always present in the phase diagram. This is due to the $\Delta_2\ne \Delta_1$ parameter value, which imbalances the energy cost for double versus single anyon occupancies. Interestingly, all  discontinuous jumps in the current are diagnosed with a level inversion, rather than any non-analyticity in the ground state. There are multiple level crossings along the $\nu$, while the level crossings along the $t$ axis for fixed $\nu$ is limited to a narrow window near  $\nu\sim 1/4$ and $\sim 3/4$ regions.

\begin{figure}[t]
    \centering    \includegraphics[width=0.95\columnwidth]{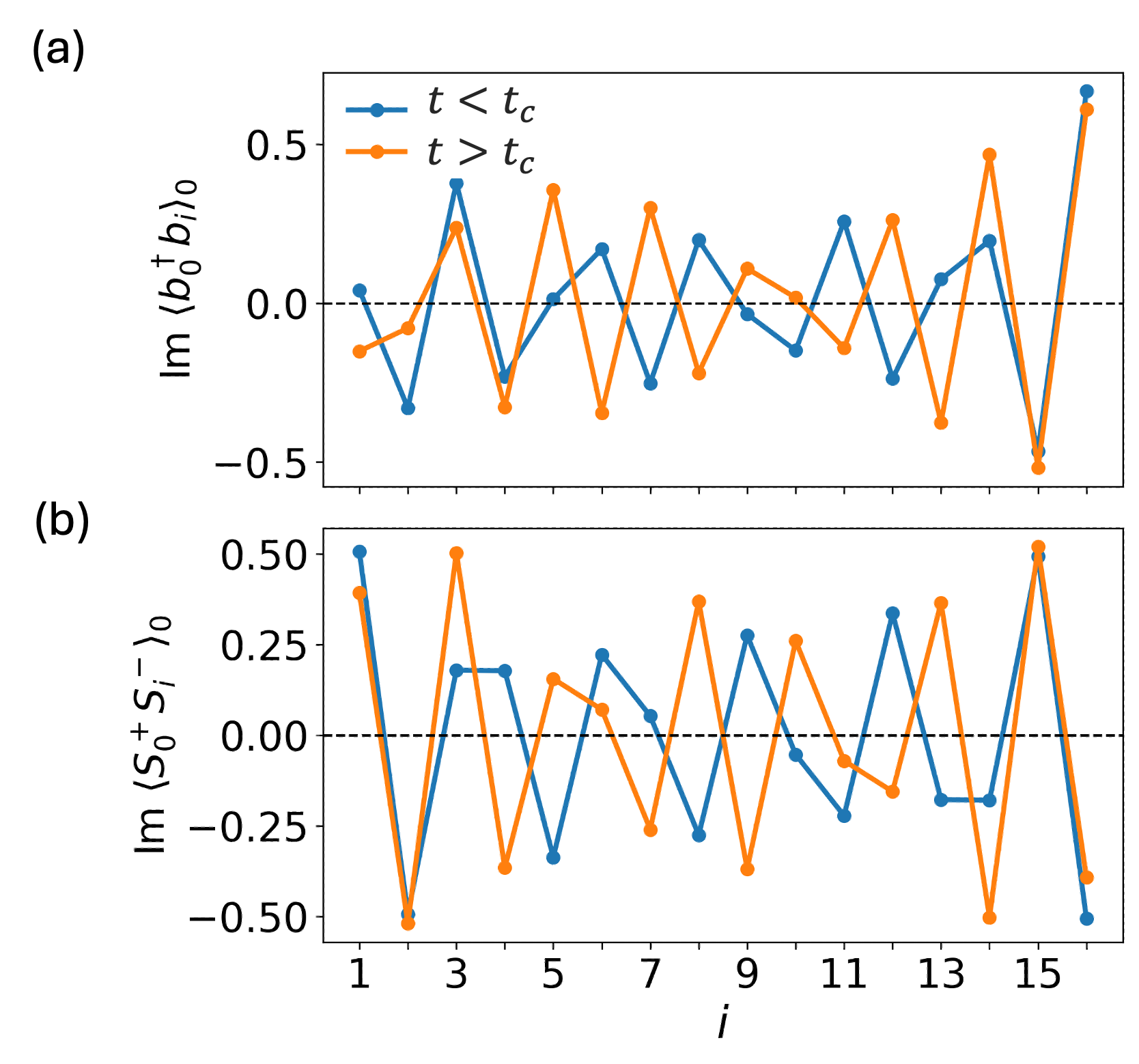}
    \caption{\textbf{Two point correlation functions.} We show the imaginary parts of the equal time correlation functions: (a) $G_{ij}$ corresponds to `Green's' function $G_{0i}$ for anyons, and (b) $\chi^{+-}_{ij}$ gives transverse spin correlation function. We set $i=0$ and plot as a function of $j\rightarrow j$. We show results for two representative values of $t$ across critical hopping $t_c=0.4$, $\nu\sim 0.6$, $\Delta_1=1$, and $\Delta_2 = 4$. For $t<t_c$, all curves corresponding to  $t \in [0.36, 0.39]$ overlap exactly. Similarly, the curves corresponding to $t\geq t_c$ for $t \in [0.40, 0.43]$ are also exactly overlapping. Both correlation functions exhibit spatial modulation, with a sudden change in amplitude and wavevector at $t_c$.
    }   \label{Fig:ancf_real_imag_gs_L12_M11M15}
\end{figure}

We calculate the equal-time two-point correlation functions between anyons, and spin operators defined as
\begin{eqnarray}
G_{ij} &=& \left\langle b^{\dagger}_jb_{i}\right\rangle = \frac{1}{2}\left\langle S_{j}^{+}e^{- i\theta\sum_{k=i}^{j-1}N_k}S_{i}^{-}\right\rangle,\nonumber\\
\chi^{+-}_{ij} &=& \left\langle S^{+}_iS^{-}_j\right\rangle = 2 \left\langle b_{i}^{\dagger}e^{- i\theta\sum_{k=i}^{j-1}N_k}b_{j}\right\rangle, \nonumber\\
\end{eqnarray}
where we adopt the convention $j>i$.
Due to PBC, we can set $i=0$ and plot these quantities as a function of $|j-i|=j$. Figs.~\ref{Fig:ancf_real_imag_gs_L12_M11M15}(a) and~\ref{Fig:ancf_real_imag_gs_L12_M11M15}(b) show the imaginary parts of $G_{ij}$ and  $\chi_{ij}^{+-}$, respectively. 

We observe periodic modulation in both correlation functions, with the same periodicity for all $t<t_c$, but changes to another fixed wavelength for $t>t_c$. In fact, we notice that the wavelength reduces by about a lattice constant above $t_c$. The result suggests that the ground states consists of a density wave like modulation in occupation charge density wave in both cases but with different wavevectors which undergoes level inversions across the phase transitions. The exact phase transition point is not numerically accessible as for any $t\rightarrow t_c+0^{\pm}$ the ground state settles to different density wave patterns. 

\section{Discussion and Conclusions}\label{Sec:Discussions}

In this manuscript, we have primarily focused on developing an algebraic formulation of second-quantized anyons, namely a Fock space construction based on a modified definition of the number operator $\hat{N}$. The key step was to 
treat the number operator $\hat{N}$ in the exponent of the graded algebra (Eq.~\eqref{Eq: Anyon algebra}), rather than by the product of creation and annihilation operators, $b^{\dagger}b$. As a result, while $\hat{N}$ remains a well-defined operator taking integer eigenvalues values corresponding to the occupancy of anyons, the operator $b^{\dagger}b$ becomes a degenerate operator (Eq.~\eqref{Eq: Anyon betan}). These features overcome a central limitation of prior lattice approaches to anyons, where exchange statistics was engineered through synthetic density‑dependent gauge fields but failed to enforce correct on-site occupancy. \cite{rabello1995gauge, rabello19961d, kundu1999exact, keilmann2011statistically,roushan2017chiral, clark2018observation, lienhard2020realization, yao2022domain, frolian2022realizing} 

Another important milestone is the exact duality between $\pi/3$ anyons and spin-1 operators (Eq.~\eqref{Eq: Anyon JW}). This duality is also formally generalized to  $\pi/N$ anyons for general $N$; however, the corresponding spin-like operators no longer obey the exact $SO(3)$ algebra, but instead satisfy an algebraic structure (see Appendix~\ref{Appendix: Anyon JW proof}). The key message is that one may start from a spin-1 lattice model $-$ which are considerably easier to manipulate both theoretically and experimentally $-$ and systematically deform them to realize anyonic statistics in tabletop experimental settings. 

We further observe rather unusual finite ground state momentum and persistent current on a ring geometry with PBC. Remarkably, the current is suppressed near half‑filling, and exhibits discontinuous jumps in both the magnitude and the direction as the system parameters are varied. Similar behavior has been reported previously in tight-binding models of anyons,~\cite{edmonds2013simulating} 
and has also been observed in experiments~\cite{roushan2017chiral}.  
The pronounced current fluctuations near such critical points suggest that noise‑based measurements could provide an effective probe for determining the fractional charge quanta of anyons.

At present, both the algebraic construction and the resulting many‑body states are restricted to one‑dimensional chains. Extending the algebraic approach to two dimensions is a natural next step but raises nontrivial questions about normal ordering and braid‑group representations in higher‑dimensional Fock spaces. Previous progress in this direction has relied on explicit two‑dimensional representations of the braid group\cite{leinaas1977} and on parastatistics constructions\cite{wang2025}. An immediate goal of future work would be to investigate whether such approaches can be consistently incorporated within the modified algebra proposed here. Finally, the spin‑based realization opens the door to device‑level explorations in platforms that already support spin‑1 physics, offering a practical pathway toward engineered anyonic matter and its applications

\begin{acknowledgments}
We thank Yuval Gefen for numerous stimulating discussions
and for a critical reading of this manuscript.
We also thank Kaden Hazzard, Rabi Narayan Mishra, Ganapathy Murthy, and Sumathi Rao for useful discussions. D.S. thanks SERB, India, for support through Project No.
JBR/2020/000043. T.D. acknowledges funding from Core Research Grant (CRG) of S.E.R.B. (CRG/2022/003412), and ANRF Advanced Research Grant (ARG) (ANRF/ARG/2025/002611/PS) and benefited from the computational resources (SERC) in
the Indian Institute of Science. P.B. thanks the Prime Minister's Research Fellowship
(PMRF) from the Government of India for financial support, and also acknowledges the technical support and computational resources (SERC) provided by the Indian Institute of Science. 
\end{acknowledgments}

\appendix

\section{Consistency Checks} \label{Appendix: Anyon consistency check}
In this section, we carry out a few consistency checks on the algebra proposed in Eq.~\eqref{Eq: Anyon algebra}. We examine the behavior of anyons on a $1D$ ring lattice under several representative exchange processes involving two anyons on a three-site lattice and a four-site lattice. In each scenario, we consider the motion of anyons to be governed by nearest-neighbor exchange phase and onsite exclusion statistics in Eq.~\eqref{Eq: Ham hopping PBC}.

\begin{figure}[th]
    \centering    \includegraphics[width=0.95\columnwidth]{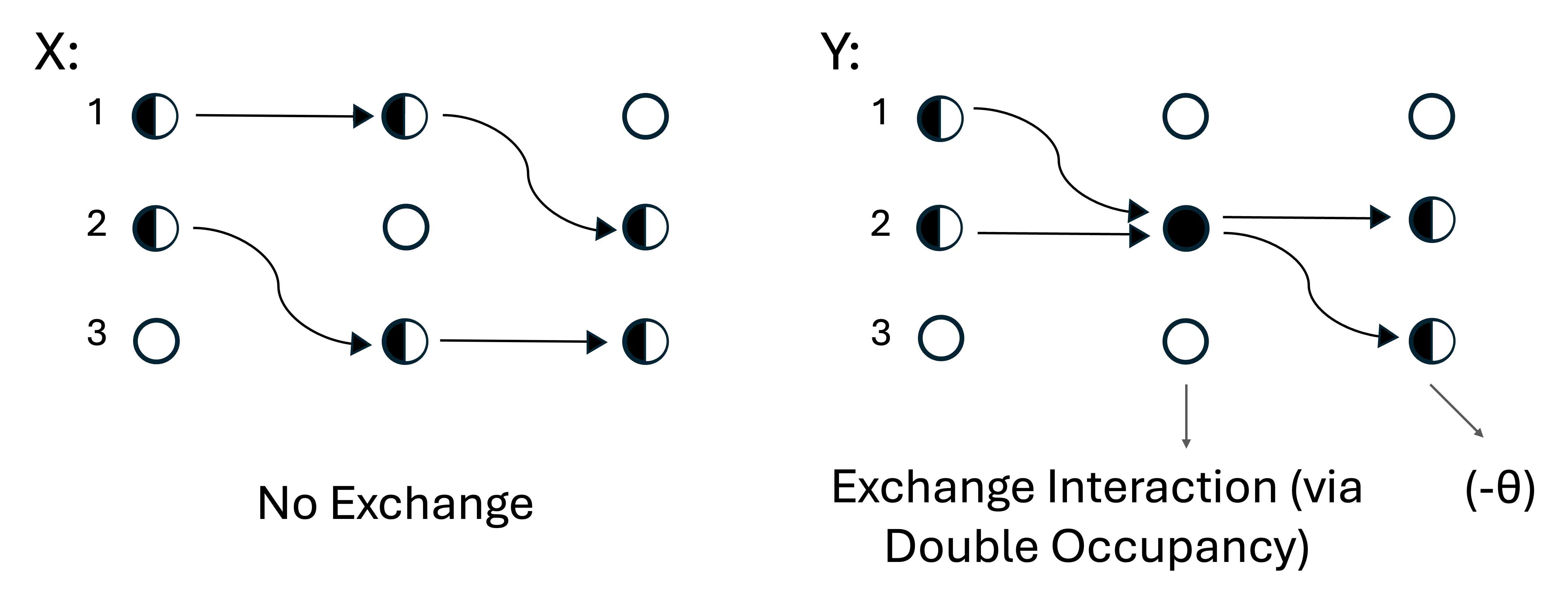}
    \caption{\textbf{Three-site two-anyon processes.} Schematic of hopping processes $X$ and $Y$ discussed in Section~\ref{Sec: Anyon 3 site 2 anyon}. The three sites are labeled $1$, $2$ and $3$. The processes are studied on a periodic lattice, i.e., sites $1$ and $3$ are connected (not shown explicitly for clarity). Unfilled, half-filled, and fully-filled circles represent unoccupied, singly occupied, and doubly occupied sites, respectively. Arrows dictate hopping between adjacent sites with hopping operators $b_i^\dagger b_{i+1}$ and $b^\dagger_{i+1} b_i$. Any nontrivial phase ($\pm\theta$) acquired during a process is indicated at the relevant steps. Process $X$ (Left panel): No exchange or double occupancy is involved here, resulting in a net phase $\theta_X=0$. Process $Y$ (right panel): It involves exchange process through double occupancy, resulting in a net phase $\theta_Y=-\theta$.
    }
    \label{Fig:sanity_checks_XY}
\end{figure}

\subsection{Three-Site Two-Anyon Processes} \label{Sec: Anyon 3 site 2 anyon}

We label the two anyon states as $(ij)$, where the first anyon occupies site $i$ and the second occupies site $j$, where $i,j\in\{1,2,3\}$. Assuming $i<j$, this state is created by $b^\dagger_i b^\dagger_j|0\rangle$. We focus on two distinct exchange processes that take the anyons initially located at sites $1$ and $2$ to the final configuration at sites $2$ and $3$ as
\begin{eqnarray}
&X: (12) \longrightarrow (13) \longrightarrow (23), \nonumber\\
&Y: (12) \longrightarrow (22) \longrightarrow (23).
\end{eqnarray}
While both processes achieve the same overall transfer of the two anyons, they do so in qualitatively different ways. Process $X$ passes through the intermediate state $(13)$, never allowing the two anyons to occupy the same site. In contrast, process $Y$ involves the intermediate state $(22)$, where both anyons temporarily occupy the same lattice site (see Fig.~\ref{Fig:sanity_checks_XY}). 

Using the algebra in Eq.~\eqref{Eq: Anyon algebra}, we find,
\begin{equation}
    X:\,\, b^\dagger_2 b^\dagger_3 |0\rangle = |23\rangle, \quad 
    Y: \,\, e^{-i\theta} b^\dagger_2 b^\dagger_3 |0\rangle = e^{-i\theta}|23\rangle.
\end{equation}
This is a notable result. In process $Y$, an extra statistical phase factor of $e^{-i\theta}$ arises from the double occupancy. In contrast, process $X$, which avoids double occupancy and thus prevents any exchange of the two anyons, carries no additional phase. This confirms the consistency of the exchange statistics of the proposed algebra.

\subsection{Four-Site Two-Anyon Processes} \label{Sec: Anyon 4 site 2 anyon}

\begin{figure*}[]
    \centering    \includegraphics[width=1\textwidth]{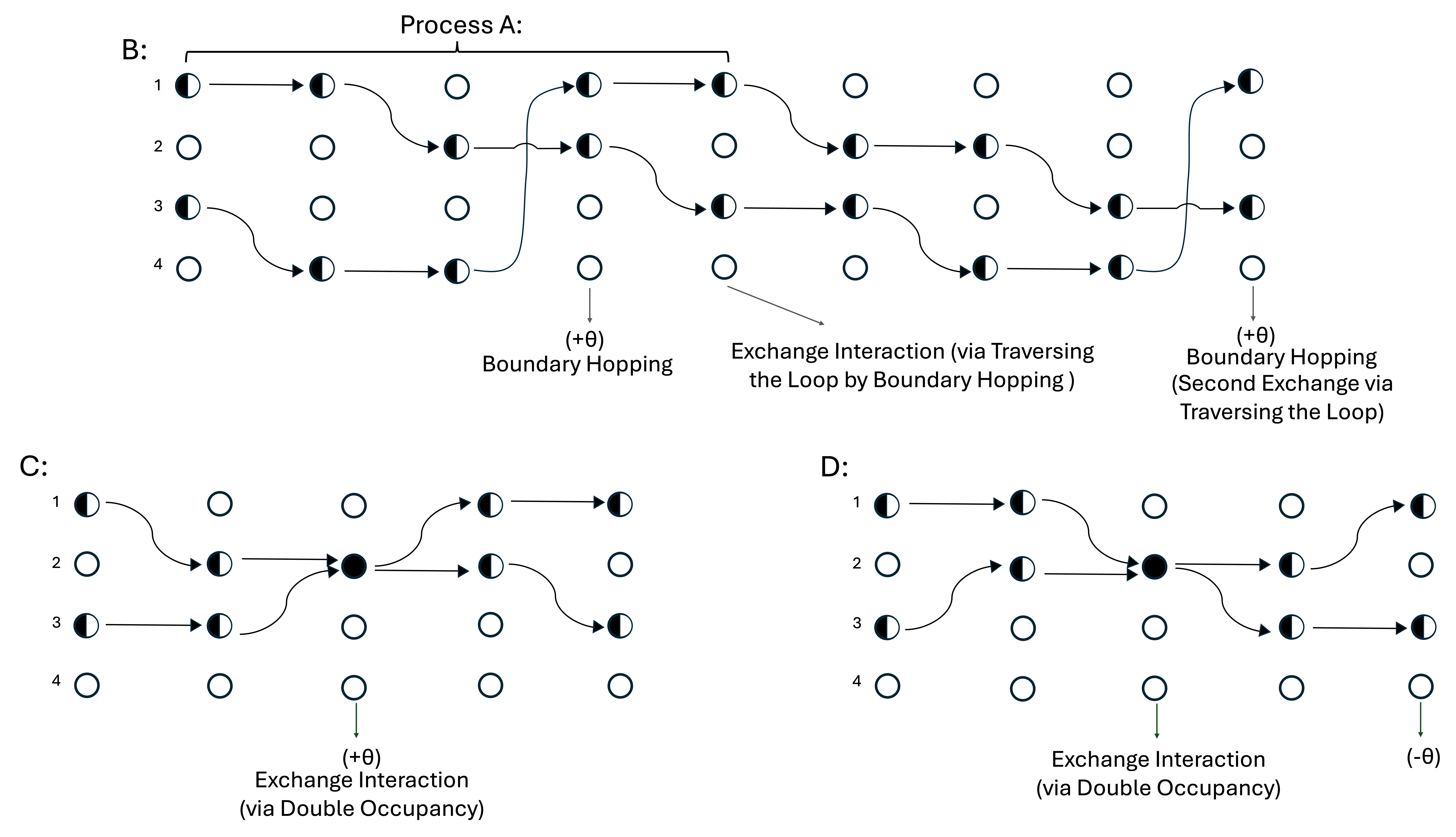}
    \caption{\textbf{Four-site two-anyon processes.} Schematic of hopping processes $A$, $B$, $C$ and $D$ discussed in Section~\ref{Sec: Anyon 4 site 2 anyon}. Four sites are labeled by $1$, $2$, $3$ and $4$. The processes are studied on a closed lattice, i.e., sites $1$ and $4$ are connected (not shown explicitly for clarity). Unfilled, half-filled, and fully-filled circles represent unoccupied, singly occupied, and doubly occupied sites, respectively. Arrows dictate hopping between adjacent sites with hopping operators $b_i^\dagger b_{i+1}$ and $b^\dagger_{i+1} b_i$. Any nontrivial phase ($\pm\theta$) acquired during a process is indicated at the relevant step. Process $A$ coincides with the first half of process $B$. Process $A$ (top panel, left sequence): Exchange of two anyons by traversing a full loop through the boundary hopping terms results in a net phase of $\theta_A=+\theta$. Process $B$ (top panel, full sequence): Two successive exchanges by traversing full loop twice gives a net phase of $\theta_B=2\theta_A$. Process $C$ (bottom left): Exchange interaction through double occupancy results inb a net phase of $\theta_C=\theta$. Process $D$ (bottom right): Exchange interaction through double occupancy, carried out in opposite direction compared to process $C$, produces a net phase $\theta_D=-\theta_C$.
    }
    \label{Fig:sanity_checks_ABCD}
\end{figure*}

Next, we examine the following four processes on a 
four-site lattice containing two anyons,
\begin{align}\label{Eq: Anyon four site process}
    & A:\, (13) \longrightarrow (14) \longrightarrow (24) \longrightarrow (21) \longrightarrow (31), \nonumber \\
    & B:\, (13) \longrightarrow (14) \longrightarrow (24) \longrightarrow (21) \longrightarrow (31) \nonumber \\
    &~~\quad \qquad  \longrightarrow (32) \longrightarrow (42) \longrightarrow (43) \longrightarrow (13), \nonumber \\
    & C:\, (13) \longrightarrow (23) \longrightarrow (22) \longrightarrow (12) \longrightarrow (13), \nonumber \\
    & D:\, (13) \longrightarrow (12) \longrightarrow (22) \longrightarrow (23) \longrightarrow (13).
\end{align}
Each of these processes involves two anyons initially separated by one lattice site. The resulting state after completing each path is either an exchange of the two anyons or a return to the original configuration (see Fig.~\ref{Fig:sanity_checks_ABCD}).

We first consider process $A$. Even though it never involves a doubly occupied state, the ring geometry of the $1D$ lattice facilitates an exchange of their positions as they traverse the lattice. Hence, we expect this path to accumulate a net exchange phase of $\theta_A = \pm \theta$, with the sign determined by the sense of rotation around the ring. In contrast, process $B$ also avoids any double occupancy but is more intricate as it returns the anyons to their original positions only after they have undergone two exchanges (in the same sense of rotation as process $A$). Consequently, process $B$ should yield a total phase $\theta_B$ that is twice that of process $A$, i.e., $\theta_B = 2\theta_A$.

Processes $C$ and $D$ are distinct from the above two processes as they both involve doubly occupied sites. This double occupancy allows the two anyons to effectively exchange at the $(22)$ step, and hence these are expected to acquire a statistical phase of $\pm \theta$. Moreover, careful inspection of their paths shows that these two processes traverse the lattice in opposite directions, so the total phases picked up must have opposite signs. Hence, their accumulated exchange phases should satisfy the relation $\theta_C = -\theta_D$.

By using the nearest neighbor hopping terms together with the anyon algebra from Eq.~\eqref{Eq: Anyon algebra}, and carefully keeping track of the operator ordering along the lattice, we obtain the following phases: $\theta_A = \theta$, $\theta_B = 2\theta$, $\theta_C = \theta$, and $\theta_D = -\theta$. These results align precisely with the qualitative behavior anticipated for each of the processes discussed above.

\section{Duality between non-local anyon operators and local ``anyon-like" operators} \label{Appendix: Anyon JW proof}

In this section, we generalize the spin-$1$ duality for $\pi/3$ anyons to the case of more general $N \neq 3$. We introduce another set of \textit{local} ``anyon-like" creation and annihilation operators $B_i^\dagger$, $B_i$, and their corresponding number operator $\hat{N}_i^B \ne B_i^{\dagger}B_i$ as 
\begin{subequations}
    \begin{align}
        & \quad B_i B_i^\dagger - e^{i\theta} B_i^\dagger B_i = e^{-i\theta \hat{N}_i^B} ,  \\
        & [B_i, B_i] = [B_i^\dagger, B_i^\dagger] = [B_i, B_i^\dagger] = 0.
    \end{align}
\end{subequations}
On the same site, the $B_i$ operators satisfy the same anyon algebra introduced in Eq.~\eqref{Eq: Anyon bbdaggerN}, except they commute on different sites like the spin or the bosonic operators. The operator $\hat{N}_i^B$ counts the number of $B_i$ type local anyons on site $i$ similar to the operator $\hat{N}_i$ in Eq.~\eqref{Eq: Anyon algebra}. Owing to the operators $b_i$ and $B_i$ sharing the same onsite commutation algebra, we get $\hat{N}_i^B = \hat{N}_i$. 

For the anyon creation and annihilation operators in Eq.~\eqref{Eq: Anyon algebra}, we note that
\begin{align}\label{Eq: Anyon Appendix JW supporting}
    b_j e^{-i\hat{N}_j\theta} &= e^{-i\theta} e^{-i\hat{N}_j\theta} b_j, \nonumber \\
    b^\dagger_j e^{-i\hat{N}_j\theta} &= e^{i\theta} e^{-i\hat{N}_j\theta} b^\dagger_j, \nonumber\\
    [\hat{N}_j,b_j^\dagger] &= b_j^\dagger.
\end{align}
Additionally, on different sites, the non-local anyon operators obey the following commutation relations, $[\hat{N}_i, \hat{N}_j] = [b_i, \hat{N}_j] = [b_i^\dagger, \hat{N}_j] = 0$ for $i\neq j$. Using these, we can relate the operators $b_i$ (non-local anyons) and $B_i$ (local anyon-like) through the following modified Jordan-Wigner transformation,
\begin{subequations}\label{Eq: Anyon anyon JW}
    \begin{align}
         b_i  &= e^{+{\rm i}\theta\sum_{j=1}^{i-1}\hat{N}_j} B_i,\\
        b^\dagger_i &= B^\dagger_i  e^{-{\rm i}\theta\sum_{j=1}^{i-1}\hat{N}_j}.
    \end{align}
\end{subequations}
For the special case of $\theta=\pi/3$, this coincides with the spin-$1$ duality presented in Eq.~\eqref{Eq: Anyon JW}. However, for $\theta = \pi/N$ where $N \ge 4$, the operators $B_i$ and $B_i^\dagger$ are not related in a linear way to the lowering and raising operators for 
any spin $S$. We find this to be
true even if we choose the spin quantum number $S = (N-1)/2$ 
for which the
number of states, $N=2S+1$, matches at each site between the
anyon and spin models.

\section{Particle Number Dependent Phase in Hopping Amplitude}\label{Appendix: Anyon tight binding hamiltonian}

Here we systematically derive the appearance of the density dependent phases $e^{\pm i (M-1)\theta/L}$ in the hopping terms in Eq.~\eqref{Eq: Ham hopping PBC}. To this end, we first examine the action of the boundary hopping term $b_1^\dagger b_L$ and the analogous bulk hopping terms $b^\dagger_{j+1} b_j$ for $j=1, \cdots, L-1$, on a general occupation number basis state. Let us first look at the action of $b^\dagger_{j} b_{j+1}$,
\begin{align}
    & b^\dagger_{j} b_{j+1} |n_1, n_2, \cdots , n_L \rangle  \nonumber \\
    & = e^{i\theta n_j}\sqrt{\beta_{n_j+1} \beta_{n_{j+1}}} |n_1, n_2, \cdots, n_j+1, n_{j+1}-1, \cdots , n_L \rangle. \label{Eq: Anyon Appendix boundary hopping gen state 2} 
\end{align}
Next, we look at the action of $b_L^\dagger b_L$,
\begin{align}\label{Eq: Anyon Appendix bulk hopping gen state}
    & b^\dagger_{L} b_{1} |n_1, n_2, \cdots , n_L \rangle  \nonumber \\
    & = e^{-i(M-1)\theta}e^{i\theta n_L}\sqrt{\beta_{n_1} \beta_{n_{L}+1}} |n_1-1, n_2, \cdots , n_L+1 \rangle. 
\end{align}
A comparison of Eqs.~\eqref{Eq: Anyon Appendix bulk hopping gen state} and ~\eqref{Eq: Anyon Appendix boundary hopping gen state 2}, with the following correspondences $j \longrightarrow L$ and $(j+1) \longrightarrow 1$, reveals that the phase acquired at the boundary hopping process differs from that of the bulk hopping process by $e^{-i(M-1)\theta}$. Consequently, to restore lattice translational invariance, an additional phase of $e^{+i(M-1)\theta}$ has to be incorporated with the boundary term $b^\dagger_L b_1$. A similar analysis can be carried out for the boundary term $b^\dagger_1 b_L$, where the required additional phase turns out to be $e^{-i(M-1)\theta}$. 

This results in the following tight-binding Hamiltonian
\begin{align}\label{Eq:TBC}
    H_{\text{hop}} &= t\sum_{j=1}^{L-1} 
    \left(b_j^\dagger b_{j+1} +b_{j+1}^\dagger b_j\right)  \nonumber \\
    &  \quad \quad + t\left(e^{i(M-1)\theta}b_L^\dagger b_1  + e^{-i(M-1)\theta}b_1^\dagger b_L \right).
\end{align}
However, this obeys a twisted boundary condition. To recover a PBC, we perform a local unitary transformation $V=\prod_{j} e^{-i (j-1) \frac{(M-1)\theta}{L} \hat{N}_{j}^{z}}$, such that
\begin{equation}\label{Eq: Anyon spin unitary trans 2}
    V b_{j}^{\dagger} V^{\dagger}=e^{- i (j-1) \frac{(M-) \theta}{L}}  b_{j}^{\dagger},
    \quad 
      V \hat{N}_{j}^{z} V^{\dagger}=\hat{N}_{j}^{z}. 
\end{equation}
This action follows from Eq.~\eqref{Eq: Anyon Appendix JW supporting}, and recovers the particle number dependent phases in the hopping term in Eq.~\eqref{Eq: Ham hopping PBC} (see Fig.~\ref{Fig:Ring hopping}).

\section{Hard-core Limit}\label{Appendix: hard core}

We consider the hard-core limit of the anyon Hamiltonian with PBC in Eq.~\eqref{Eq: Ham hopping PBC}. Specifically, we disallow any doubly occupied site, i.e., locally the anyon state $|N_i=2\rangle$, or equivalently in spin-$1$ language $|S^z_i=+1\rangle$ states are not allowed. The hardcore constraint is implemented by associating an infinite energy cost to doubly occupancy, i.e., $\Delta_2\rightarrow \infty$. This suggests a mapping from spin-$1$ to spin-$1/2$ as follows,
\begin{equation}
    S^{\pm}_j=\sqrt{2}\sigma_j^\pm, \quad S^z_j=\frac{\sigma^z_j-1}{2},
\end{equation}
where the $S_j^a$'s represent spin spin-$1$ operators and $\sigma_j^a$ represent Pauli matrices.

The spin-$1$ Hamiltonian without the unitary gauge transformation implemented in Sec.~\ref{Sec:lattice_model} (see the discussion following Eq.~\eqref{Eq: Anyon hamiltonian spin periodic}) is,
\begin{align}\label{Eq: Anyon hamiltonian spin twisted}
    H &= \frac{\tilde{t}}{2} e^{i\theta} ~\sum_{j=1}^{L-1}  
      ~(S_{j}^{+} e^{{\rm i}S_j^z\theta} S_{j+1}^{-} ~+~ {\rm H.c.}) \nonumber \\
    & \quad +\frac{\tilde{t}}{2}  e^{i\theta} e^{-i(M-1)\theta}
      S_{L}^{+} e^{{\rm i}S_L^z\theta} S_{1}^{-} + \text{H.c.}
     \nonumber \\
    & \quad + \sum_{j=1}^L ~\left[ V (S^z_j)^2 + h S^z_j   \right]+L\Delta_1, 
\end{align}
where $\tilde{t}=t e^{i(M-1)\theta/L}$, and H.c. denotes the Hermitian conjugate of the hopping terms. In terms of spin $1/2$ operators, the Hamiltonian becomes
\begin{align}
    H &= (\tilde{t} \sum_{j=1}^{L-1}  
      \sigma_{j}^{+}  \sigma_{j+1}^{-}
      + \tilde{t}  e^{-i(M-1)\theta} 
      \sigma_{L}^{+}  \sigma_{1}^{-}) +\text{H.c.}
      \nonumber \\
       & \qquad + \frac{\Delta_1}{2} \sum_{j=1}^L (\sigma^z_j+\mathbb{I}).
\end{align}
Next, we implement the Jordan-Wigner transformation from spin-$1/2$ operators to spinless fermion creation and annihilation operators~\cite{Jordan1928} given as
\begin{subequations}\label{Eq: fermion JW}
    \begin{align}
        \sigma_i^z &=2 c_j^\dagger c_j-1, \\
         \sigma_i^- &=  e^{{\rm i}\pi\sum_{j=1}^{i-1}c^\dagger_j c_j} c_i,\\
        \sigma^+_i &=  c^\dagger_i  e^{{\rm i}\pi\sum_{j=1}^{i-1}c^\dagger_j c_j}.
    \end{align}
\end{subequations}
This maps a site with $\sigma^z=-1$ to an empty fermion site and that with $\sigma^z=+1$ to a filled fermion site. Under this transformation, the anyon Hamiltonian reduces to a system of non-interacting fermions on a $1D$ lattice under twisted boundary conditions given as,
\begin{align}
    H &= \tilde{t} \sum_{j=1}^{L-1}~  
      (c^\dagger_j c_{j+1} ~+~ {\rm H.c.}) \nonumber \\
      & ~~~+~ \tilde{t} (-1)^{M-1} e^{-i(M-1)\theta} 
      c^\dagger_L c_1 ~+~ \text{H.c.}
      \nonumber \\
       & ~~~ + ~\Delta_1 ~\sum_{j=1}^L c^\dagger_j c_{j}.
\end{align}
After a Fourier transformation of the fermionic modes, we obtain
\begin{align}
    H &=\sum_k E(K) c^\dagger_K c_K +\Delta_1 M, \nonumber \\
    \text{where}~~ E(K) &= 2t \cos \left[ K+ (M-1)\frac{\theta}{L} \right], 
\end{align}
and the momenta $K$ are quantized as,
\begin{align}\label{Eq: k hb plus anoyn}
    K &= \frac{2\pi}{L}k +(M-1)\frac{\pi}{L} -(M-1)\frac{\theta}{L}, \quad k\in \mathbb{Z}. \nonumber \\
    & = K_{\text{HB}} -(M-1)\frac{\theta}{L},
\end{align}
where $K_{\text{HB}}$ represents the momenta of the hard-core bosons on a $1$D ring lattice with the many body wavefunction written as a single Slater determinant. Here exchange of bosons is only allowed through cyclic permutations as double occupancy is disallowed under 
the hard-core constraint. The expression of $K_{\text{HB}}$ in the above equation coincides with that of the Thouless and Li treatment of hard-core bosons in~\cite{Li1987} with the excluded volume $d=0$. Hence the wavefunction of $M$ anyons in the hard-core limit acquires a phase of $(M-1)\theta$ under a cyclic permutation of $M$ anyons.

\section{System size dependence}\label{Appendix: System size invariance}

\begin{figure*}[t]
    \centering    \includegraphics[width=0.95\textwidth]{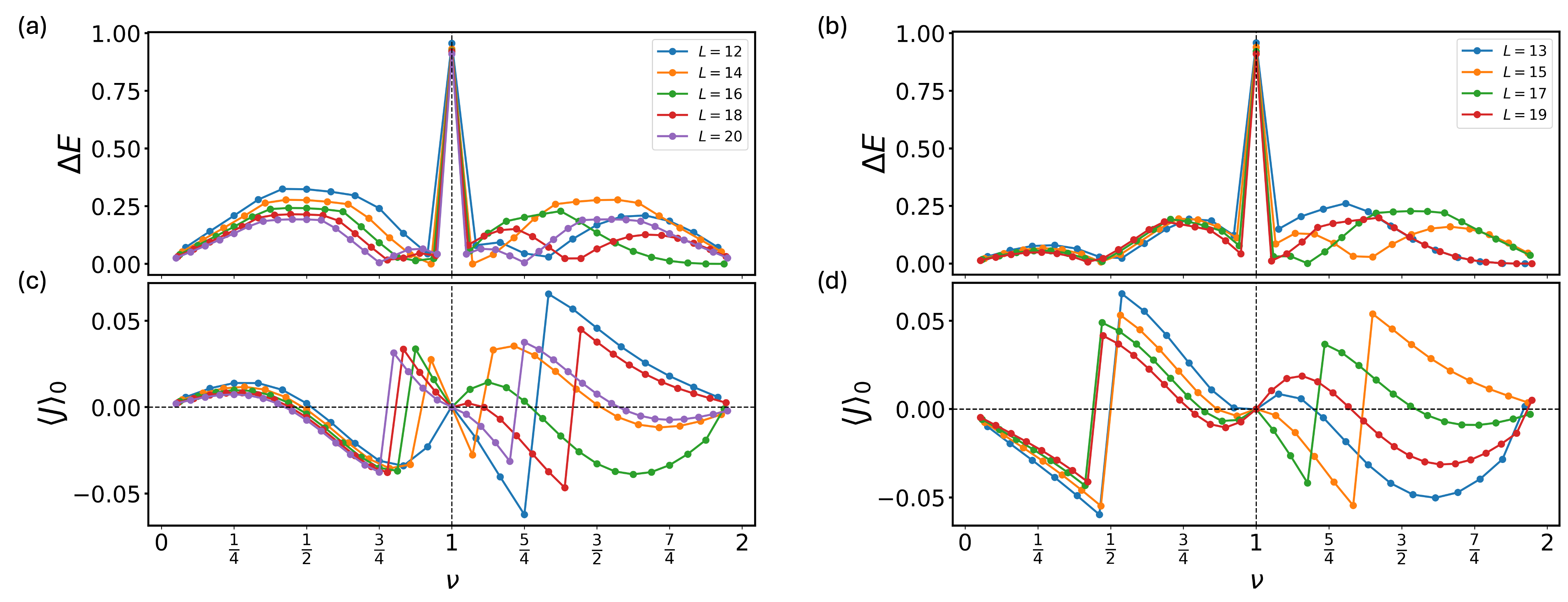}
    \caption{\textbf{System size dependence.} Left panel: (a) Energy gap $\Delta E=E_1-E_0$ between the first excited state and the ground state, and (c) the ground state persistent current $\langle J \rangle_0$, plotted as a function of anyon number density $\nu$ for even number of sites $L=12,14,16,18,20$, with fixed hopping amplitude $t=0.4$, and onsite interaction defined by $\Delta_1=1$, $\Delta_2=4$ (i.e., $U=2>0$). The vertical dashed line marks half-filling ($M = L$).
    Right panel: The same quantities plotted for odd number of sites $L=13,15,17,19$. 
    Dots represent numerical data, while lines are guides to the eye.
    }
    \label{Fig:PC 1st 2nd jump eg L15M18}
\end{figure*}

In this section, we demonstrate that the results discussed in the main text remain qualitatively invariant with varying system sizes. In Fig.~\ref{Fig:PC 1st 2nd jump eg L15M18}, we plot the energy gap $\Delta E$ and the ground state current $\langle J \rangle_0$ for varying system sizes, for fixed $t=0.4$, $\Delta_1=1$, and $\Delta_2=4$. Left/right panel of the figure displays the variation for even/odd number of sites. We note that the correspondence between the current sign reversal and the suppression in the energy gap discussed in the main text, holds for varying system sizes.

From Fig.~\ref{Fig:PC 1st 2nd jump eg L15M18}, we also note a qualitative difference between the odd and even system sizes. Specifically, for $\nu<1$, the even system sizes (left panel) exhibit current sign reversal for $\nu\sim3/4$ whereas the odd system sizes (right panel) exhibit the same for $\nu\sim 1/2$. The investigation of the underlying physical behavior leading to the distinction between odd and even system sizes will be a part of the future studies.

\section{Persistent current for $\pi/3$ anyons}\label{Appendix: Anyon PC}

In this section, we derive the expression for the persistent current operator in Eq.~\eqref{Eq: Anyon pc bbdagger}. For that, we use the Heisenberg equation of motion, $\dot{\hat{N}}_j =i[H,\hat{N}_j]$, along with the continuity equation on a lattice, $\dot{\hat{N}}_j + J_{j+1/2} - J_{j-1/2} = 0$. The symbol $~\dot{}~$ denotes
derivative with respect to time, and $J_{j+1/2}$ represents the current operator through the bond connecting sites $j$ and $j+1$. We first evaluate $\dot{\hat{N}}_j$ using the Heisenberg equation of motion,
\begin{align} \label{Eq: H Nj Com}
    \dot{\hat{N}}_j
    &= - it ~[ (e^{i(M-1)\theta/L} b_{j}^\dagger b_{j+1} -e^{-i(M-1)\theta/L} b_{j+1}^\dagger b_{j}) \nonumber \\
    & ~~~~~~~~~- (e^{i(M-1)\theta/L}b_{j-1}^\dagger b_{j} - e^{-i(M-1)\theta/L}b_{j}^\dagger b_{j-1}) ], \nonumber \\
\end{align}
where we have used the following commutation relations valid for $\pi/3$ anyons, $[b_j,N_j]=b_j$, $[b_j^\dagger, N_j]=-b_j^\dagger$, and $[N_i,N_j] = [b_i,N_j] = [b_i^\dagger, N_j] = 0$ for $i \neq j$. 

Next, using the continuity equation, the anyonic current operator in Eq.~\eqref{Eq: Anyon pc bbdagger} can be identified straightforwardly. Note that Eq.~\eqref{Eq: Anyon pc bbdagger} corresponds to the average current operator over the $L$ bonds. Owing to the PBC, this coincides with the current operator through a chosen bond.

\bibliography{biblio}

\end{document}